\documentclass{aa}  

\usepackage{graphicx}
\usepackage[varg]{txfonts}
\usepackage[]{natbib} 
\begin{document} 

\title{Exploring extreme conditions for star formation: a deep search for molecular gas in the Leo ring}

\author{Edvige Corbelli \inst{1}
\and David Thilker\inst{2}
\and Filippo Mannucci\inst{1}
\and Giovanni Cresci\inst{1} 
  }

\institute {INAF-Osservatorio di Arcetri, Largo E. Fermi 5, 50125 Firenze, Italy
  \  \email{edvige.corbelli@inaf.it}
   \and Department of Physics and Astronomy, The Johns Hopkins University, Baltimore, MD, USA
 }

\date{Received ....; accepted ......}

 \abstract
  {}
{ We carry out  sensitive searches for the $^{12}$CO J=1-0 and J=2-1 lines in the giant extragalactic HI ring in Leo to investigate the star formation process within environments where gas metallicities are close to solar but physical conditions are different than those typical of  bright galaxy disks. Our aim is to check the range of validity of known scaling relations. }  
{ We use the IRAM-30m telescope to observe  eleven regions  close to HI gas peaks or where sparse young massive stars have been found. For all pointed observations we reached a spectral noise between 1 and 5~mK  for at least one observed frequencies  at  2~km~s$^{-1}$ spectral resolution.
}
 { We  marginally detect two  $^{12}$CO J=1-0 lines in the star forming region Clump~1 of the Leo ring, whose radial velocities are consistent with those of H$\alpha$ lines but line widths are much smaller than observed for virialized molecular clouds of similar mass in galaxies. 
The low signal-to-noise ratio, the small line widths and the extremely low number densities  inferred by virialized cloud models suggest that a more standard population of  molecular clouds, still undetected, might be in place.  Using upper limits to the CO lines,
the most sensitive pointed observations show that the molecular gas mass surface density is lower than expected from  the extrapolation of the molecular Kennicutt-Schmidt relation established in the disk of galaxies. The sparse stellar population in the ring, possibly forming ultra diffuse dwarf galaxies, might then be the result of a short molecular gas depletion time in this extreme environment.}
 {}


\maketitle

\section{Introduction}

Although our understanding of the basic physical processes leading to star formation  is based on high resolution Galactic studies \citep[e.g.][]{2019A&A...629L...4A}, nearby galaxies have often complemented  Milky Way studies offering  a variety of  interstellar medium  (ISM) conditions to be examined   \citep{2012A&A...542A..32C,2013AJ....146...19L,2015ApJ...814L..30E,2017A&A...601A.146C,2021ApJS..257...43L}. Extragalactic surveys have widened our knowledge  of the star formation process by inspecting how this changes as galaxies evolve through cosmic time, as galaxy mass, morphology and metallicity change \citep[e.g.][]{2012ARA&A..50..531K,2012ApJ...745...69K, 2020ARA&A..58..157T,2021A&A...649A..39L}. Dwarf galaxies and outer disks of spiral galaxies have been targets for  investigating star formation laws  in low density regions \citep{2007ApJS..173..538T,2013MNRAS.436.2747K,2015ApJ...805..145E,2016MNRAS.455.1807W,2018MNRAS.480..210L,2019A&A...622A.171C}. Key questions are the formation of the molecular phase and its tracers, as well as  the scaling relation of the star formation rate with the surface or volume gas density, such as the Kennicutt-Schmidt (K-S) law \citep{2010AJ....140.1194B,2012ApJ...759....9K,2019ApJ...872...16D,2020A&A...644A.125B}. The low metallicities, however,  have often limited the interpretation of results, especially those concerning the formation and evolution of molecular clouds, building blocks for the star formation process and  traced via CO lines  \citep{2013ARA&A..51..207B,2015A&A...583A.114H,2022AJ....163..141A}. 

Very few studies have been dedicated to the formation of stars at yet another extreme: metal rich low density regions such as intergalactic neutral clouds  of tidal origin. Most studied cases are relative to shocked gas recently removed from the disk of galaxies through tidal encounters where gas and star formation densities are high \citep{2000Natur.403..867B,2004A&A...426..471L,2021A&A...645A..97Q}. The paucity of more quiescent cold clouds in the Local Universe, with only occasional formation of stars, surely limits the exploration of the slow star formation  process in these metal rich but low density environments. However, these clouds offer the unique opportunity to test star formation laws at extremely low rates where  gas physical conditions are different than in disk galaxies (with no rotational support and no stellar disk driving supersonic turbulence).  Stochasticity at the upper end of the stellar initial mass function  (hereafter IMF) and  density dependencies of the CO-to-H$_2$ conversion factor   \citep{2002ApJ...566..818L,2014MNRAS.444.3275D,2022ApJ...931...28H} requires some caution in interpreting the results, nevertheless understanding star formation in extragalactic clouds of tidal origin is very relevant  because we might capture the slow building up process of  diffuse dwarf galaxy formation \citep{2021ApJ...923L..21P}.

The serendipitous discovery of an optically dark HI cloud in the M96 group  \citep{1983ApJ...273L...1S} has since then triggered a lot of discussion on the origin
and survival of the most extended and massive HI structure in the local intergalactic medium. With an
extension of about 200 kpc and a neutral gas mass M$_{HI}=2\times10^9$~M$_\odot$, the cloud has a ring-like shape, and it
is known also as the Leo ring which might orbit the galaxies M105 and NGC3384 with a 4-Gyr period \citep{1985ApJ...288L..33S}. The ring is more than 3 Holmberg radii distant from any known galaxy and significantly larger than other collisional ring known. The lack of a significant intragroup diffuse
starlight and of an extended optical counterpart of the ring  \citep{1985AJ.....90..450P,1985MNRAS.213..111K,2014ApJ...791...38W} 
supported the ring primordial origin hypothesis  \citep{1989AJ.....97..666S, 2003ApJ...591..185S}.  Intermediate resolution HI maps (obtained with the Very Large Array, hereafter VLA, with an effective beam of 45\arcsec\ ) revealed the presence of gas clumps \citep{1986AJ.....91...13S} with column densities of order 10$^{21}$~cm$^{-2}$ with a few of these associated
to faint ultraviolet (hereafter UV) emitting regions, indicative of recent localized events of star formation \citep{2009Natur.457..990T}. 
Spectroscopic data, acquired recently for 2 major HI clumps in the Leo ring (Clump1 and Clump2E) have recently given new insights to the Leo cloud mystery revealing the presence of  ionized gas and metal lines compatible with gas metal abundances close to solar \citep{2021ApJ...908L..39C} thus finally unambiguously unveiling the ring tidal origin. Numerical simulations have provided some evidence that the gas might have been stripped from a low luminosity galaxy about 1 Gyr ago or during a galaxy-galaxy head-on collision \citep{1985ApJ...288..535R, 2005MNRAS.357L..21B,2010ApJ...717L.143M}. 

The gas in the Leo ring, pre-enriched in a galactic disk and tidally stripped, did not manage to form stars very efficiently in intergalactic space.  There has been no confirmed diffuse H$\alpha$ emission   \citep{1986ApJ...309L...9R,1995ApJ...450L..45D}  or CO detection from a pervasive population of giant molecular complexes \citep{1989AJ.....97..666S}. However, deep optical imaging \citep{2009AJ....138..338S, 2015PKAS...30..517K,2018A&A...615A.105M,2018ApJ...863L...7M} reported diffuse faint dwarf galaxies close to the ring which, together with the possibility of clump rotation inferred from the ring HI mapping \citep{1986AJ.....91...13S}, suggests  a slow in situ formation of dwarf galaxies. The H$\alpha$ luminosities of  the recently discovered ionised nebulae,  together with archival Hubble Space Telescope images, indicate unambiguously that  in fact sporadic star formation episodes are taking place in the ring during the last 10 million years  \citep{2021A&A...651A..77C}.  These localised events of star formation happens close to HI gas peaks at an average rate  of  0.0004 M$_\odot$~yr$^{-1}$~kpc$^{-2}$.  The presence of stars more massive than a B0-type and the spatial distribution of stellar cluster ages suggest that these events might have been  triggered by feedback from a previous generation of stars.

 It is interesting to investigate  gas cooling and fragmentation driving these events of star formation, whether they produce molecular clouds and gas clumps with physical characteristics similar to those in disk galaxies, such as the  size-linewidth relation,  mass surface densities, star formation efficiencies and timescales  \citep{2013ARA&A..51..207B,2013AJ....146...19L}. The inferred high metal abundances in the Leo ring make it an ideal candidate for
investigating if and how stars are formed out of molecular hydrogen and if this can be traced via CO
emission in a low density environment. The formation of stars and the molecule formation balance is still unexplored  in a simplified environment   with a lower radiation field, lower gravity (no pervasive stellar population or dark matter).

In this paper we present  the results of IRAM-30m deep pointed observations searching for the $^{12}$CO J=1-0 and J=2-1 lines in gas clumps and star forming regions of the Leo ring. Previous CO searches  at the location of a few HI peaks  have been carried out using the FCRAO telescope \citep{1989AJ.....97..666S} providing upper limits of only 0.8~K~km~s$^{-1}$.
These measures exclude that  HI clumps are close analogue of dwarf galaxies with a large molecular-to-atomic gas ratio \citep{1987ApJ...322..681T, 2005A&A...438..855I}  but leave open the possibility of a more normal molecular-to-atomic gas ratio, close to a few percent.  
Moreover, any CO detections can help understanding  how star formation propagates in such environment.

The paper plan is as follows: in Section~2 we describe the observations and the data analysis, in Section~3 we discuss the  marginally detected lines and the possibility that these are due to spurious noise. In this last case we consider the undetected population of clouds compatible with the upper limits to CO lines and  relate the molecular gas surface density to the star formation rate density in Section~4. Here we check the consistency of the Leo ring data with the scaling relation observed in galactic disks and outer regions of galaxies. Section~5 summarizes and concludes. Throughout this paper we assume that the distance to the Leo ring is 10~Mpc i.e. 1~arcsec corresponds to a scale of 48.5~pc.

\section{Observations of $^{12}$CO lines}

In this Section we describe pointed observations of gas clumps and star forming regions of the Leo ring with the IRAM-30m telescope at the frequencies of  $^{12}$CO J=1-0 and J=2-1 lines. The antenna beam, with a full-width at half maximum (hereafter FWHM) of 21.4~arcsec at 115~GHz and of 10.7~arcsec at 230~GHz, is  sufficiently large to cover each HII region  (2-3 arcsec or about 100-150 pc in size) and its surroundings in just one pointing. The possibility of simultaneously observing two CO lines at different frequencies and with different beam sizes can possibly constrain line excitation mechanism and the CO extent.  

\begin{figure}
\includegraphics [width=9.0 cm, angle=0 ]{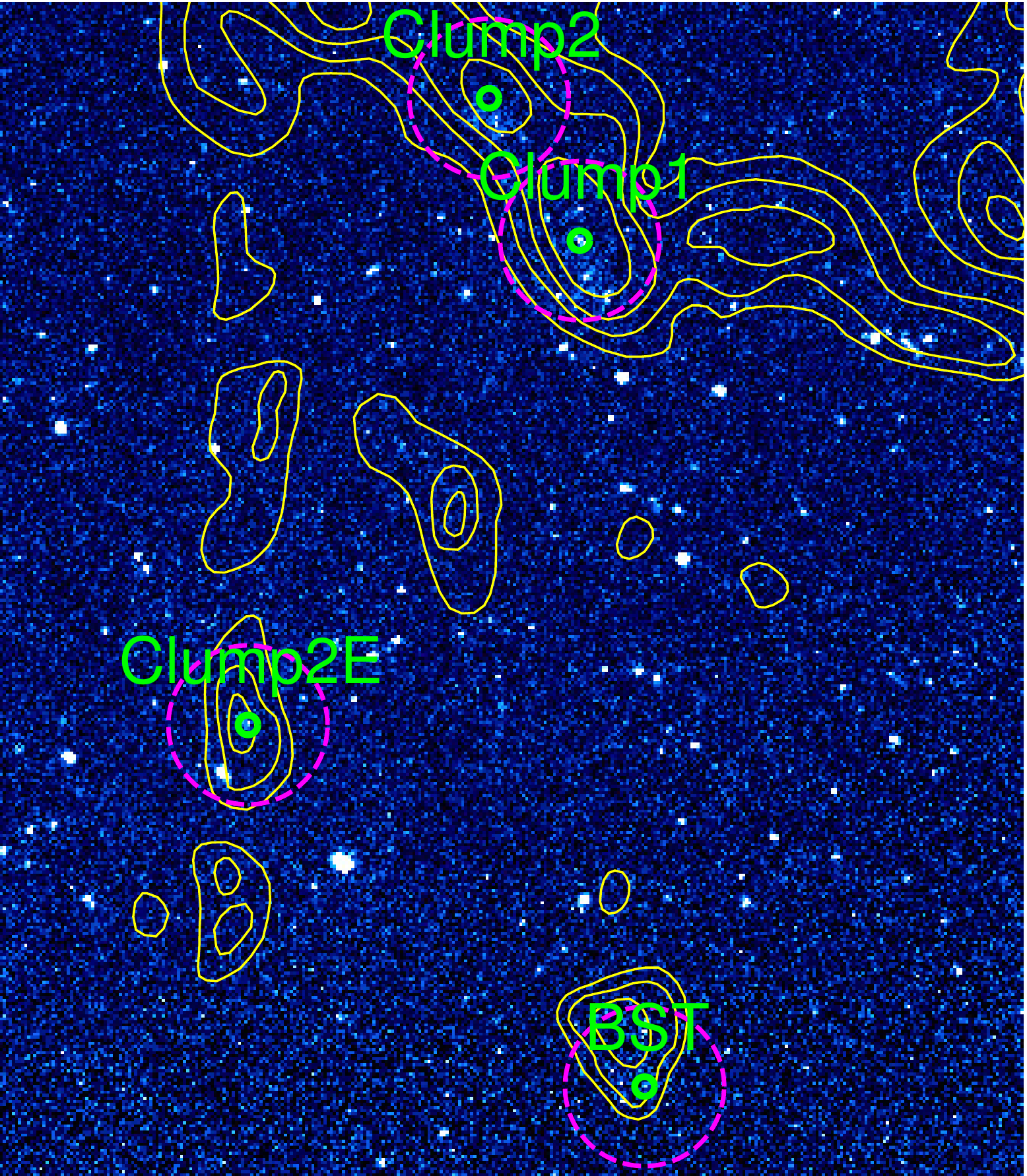}
 \caption{The four HI clumps in the Leo ring where deep searches for the CO J=1-0 and J=2-1 lines have been carried out with the IRAM-30m telescope are  marked with green circles. The circle size is  the FWHM of the 115~GHz beam (22~arcsec). More than one location has been observed in Clump1 and Clump2E where H$\alpha$ emission has been recently detected (see Figure 2 and 3). The yellow contours indicate the column density of the HI gas in the giant Leo ring as observed with the VLA (45~arcsec beam) by Schneider et al. (1986). The magenta dashed  circles indicate the location of the wobble OFF positions. The background image is the GALEX-far UV image.
}
\label{fig_hi}
\end{figure} 

We have carried out  IRAM-30m pointed observations in 4 selected areas during 5 observing periods: from November 2020 to January 2022. We observed all positions using the wobbler switching mode, setting the wobble throw at 90~arcsec. The use of the wobble produces spectra with better baselines with respect to the standard position switching ON-OFF observing mode. This becomes  very relevant  when averaging several scans i.e. for long observing periods. The locations in the ring of the four HI gas clumps observed are shown in Figure~1, where we highlight in green the FWHM of the IRAM beam at 115~GHz and the indicative locations of the wobble throw.  The wobble throw encompasses  gaseous regions of the ring and this is of concern for a possible evaluation of CO diffuse emission.  Given the localised and rare event of star formation in the ring and the ring distance,  it however unlikely that the wobbler throw  intersects  a giant molecular clouds at the same velocity as that on source, since the wobble throw is at a projected distance of more than 4~kpc from source and the expected signal is only a few km~s$^{-1}$ wide.  The FTS backend with a spectral resolution of 195 kHz was used, corresponding to channel widths of 0.5~km~s$^{-1}$ at 115~GHz and 0.25~km~s$^{-1}$ at 230~GHz. The VESPA backend system with 78.1~kHz resolution (0.2~km~s$^{-1}$) and the WILMA backend system with 10~km~s$^{-1}$ resolution were also used, but the noise level was higher than for the FTS backend when data were smoothed at the same spectral resolution, therefore we only discuss the FTS data. 

\begin{table}
\caption{Coordinates of selected regions of the Leo ring and the rms $\sigma$ of CO spectra sampled at 2~km~s$^{-1}$ channel width.} 
\centering                                       
\begin{tabular}{c c c c c }           
\hline\hline 
 Region & RA & DEC &  $\sigma_{1-0}$  &  $\sigma_{2-1}$   \\  
 \hline\hline
              &        &         &   mK     &    mK    \\
 \hline\hline 
 C1a  &   10:47:47.9 & 12:11:32.0  &  1.1 & 1.0     \\
 C1b  &   10:47:47.4 & 12:11:27.7   & 2.8  &  2.6   \\
 C1an&   10:47:47.9 & 12:11:40.0   &  5.0 &  6.6  \\
 C1ase& 10:47:48.4 & 12:11:23.1   & 4.1 & 4.0   \\
 C1c  &   10:47:46.0 & 12:11:08.6   & 4.4  &  4.1 \\
 C2Ea&  10:48:13.5 & 12:02:24.3  &  3.7 & 3.3    \\
 C2Eb&  10:48:14.1& 12:02:32.5   & 4.2 & 3.7 \\
 C2Ef &  10:48:14.5 & 12:02:20.1   & 4.3  & 3.9 \\
 BSTd &  10:47:42.9 & 12:11:23.1   & 1.5  & 1.5 \\
 Cl1    &  10:47:46.8 & 12:11:11.0   &  3.0 & 3.2 \\
 Cl2    &  10:47:54.9 & 12:14:13.0   &  4.4 &  6.0\\
 C1ab &  stacked   & stacked & 0.99 & 0.84 \\
 C1cl  & stacked  &  stacked & 2.5 & 2.7 \\
 C2E  & stacked  &  stacked & 2.2 & 1.9 \\
\hline\hline 
\end{tabular}
\label{lines}
\tablefoot{Units are  main beam temperatures.}
\end{table}

For every pointed observation our aim was to reach a spectral noise $\sigma\le 5$~mK in the averaged spectra for one of both CO lines when smoothed at a spectral resolution $\delta$V=2~km~s$^{-1}$. We list in Table~1 the coordinates of the 11 regions where we achieve this goal;  deep pointed observations have been carried out for every position during at least two distinct runs. The final rms in the CO J=1-0 spectra ranges between 1.1~mK in C1a and 5~mK in C1an i.e. the spectral sensitivity is between 27 and 6 times better than previous observations carried out by \citep{1989AJ.....97..666S}. In addition the large FCRAO beam is more strongly affected by beam dilution having a four times larger beam area. 

In Figure~2 and Figure~3 we zoom in on the regions of Clump~1 and Clump~2E where more than one  position has been observed. We mark  the observed positions with a continuous and dashed red circle whose size is the  FWHM of the IRAM beam at 115 and 230~GHz respectively. In the background we display the location of the H$\alpha$  counterparts observed with MUSE at VLT  and the far-UV (hereafter FUV) counterparts  detected by the GALEX satellite.  In Table~1 we also show the rms of stacked spectra, obtained by averaging spectra at different but spatially close locations. These are  C1ab, resulting from  C1a, C1b, C1an, C1ase averaging, C1cl, resulting from  Cl1 and C1c averaging, and C2E, resulting from C2Ea, C2EB and C2Ef averaging. Clump~2 and BST have just one pointed observation: for Clump~2 the observed position is that of the nominal HI peak  \citep{1986AJ.....91...13S}, while for BST we selected BSTD, the location of a faint FUV region with no optical counterpart (to avoid contamination by background galaxies). 

\begin{figure*}
\includegraphics [width=18.0 cm, angle=0 ]{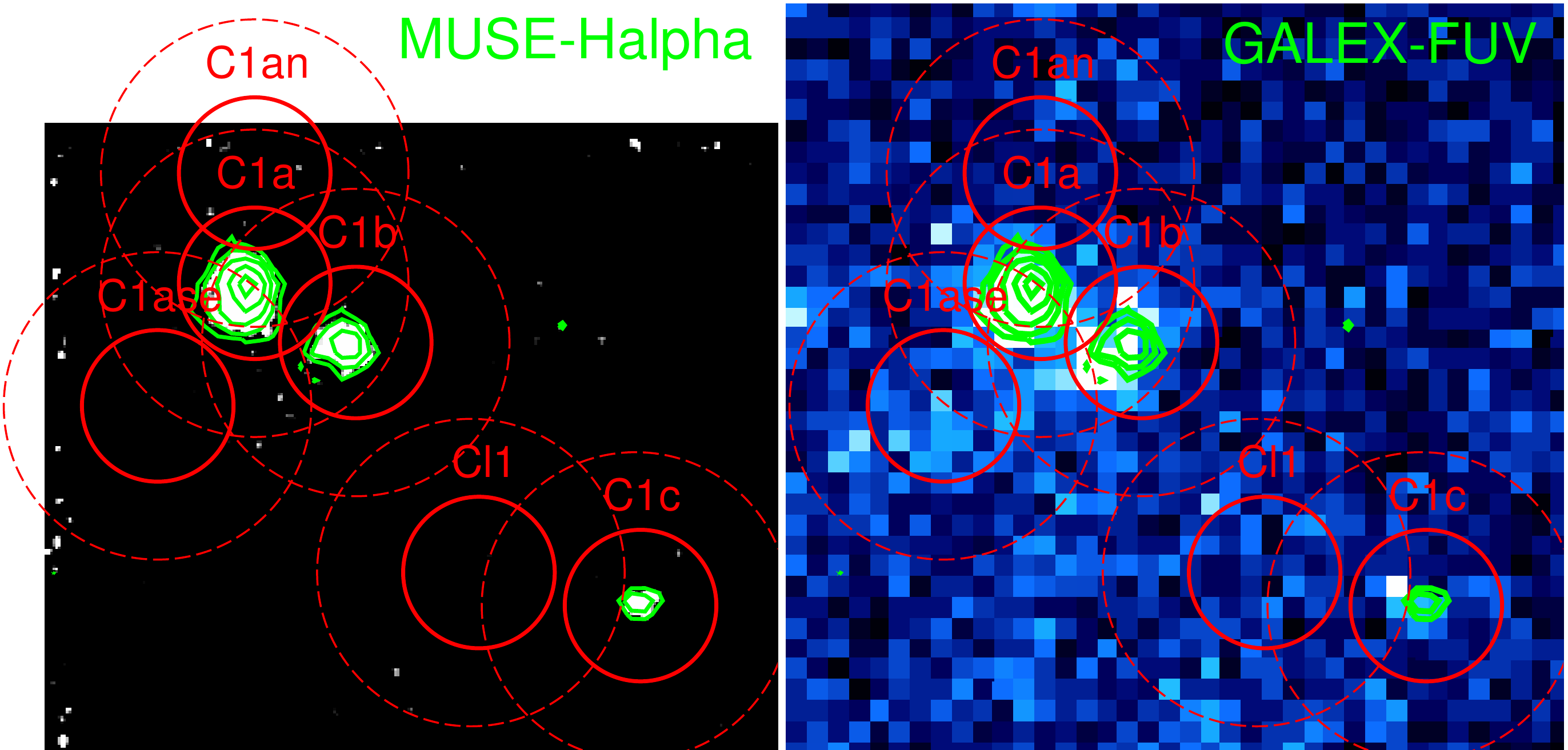}
 \caption{Contours of the H$\alpha$ emission of the three ionized regions in Clump1 (C1a, C1b, and C1c) are overlaid in green on the MUSE-H$\alpha$ image in the left panel and on the GALEX-FUV continuum image in the right panel. Contour levels are: 1.3,2.5,5,10,15$\times$10$^{-20}$~erg~s$^{-1}$~cm$^{-2}$ per pixel (0.2\arcsec ).  Red continuous lines indicate the size of the  230~GHz beam FWHM at the locations observed in Clump1 with the IRAM-30m telescope. Dashed red lines show the corresponding FWHM of the 115~GHz  beam. Observed position labels are placed  just outside the continuous red line circles.  
}
\label{fig_c1}
\end{figure*} 

\begin{figure*}
\includegraphics [width=18.0 cm, angle=0 ]{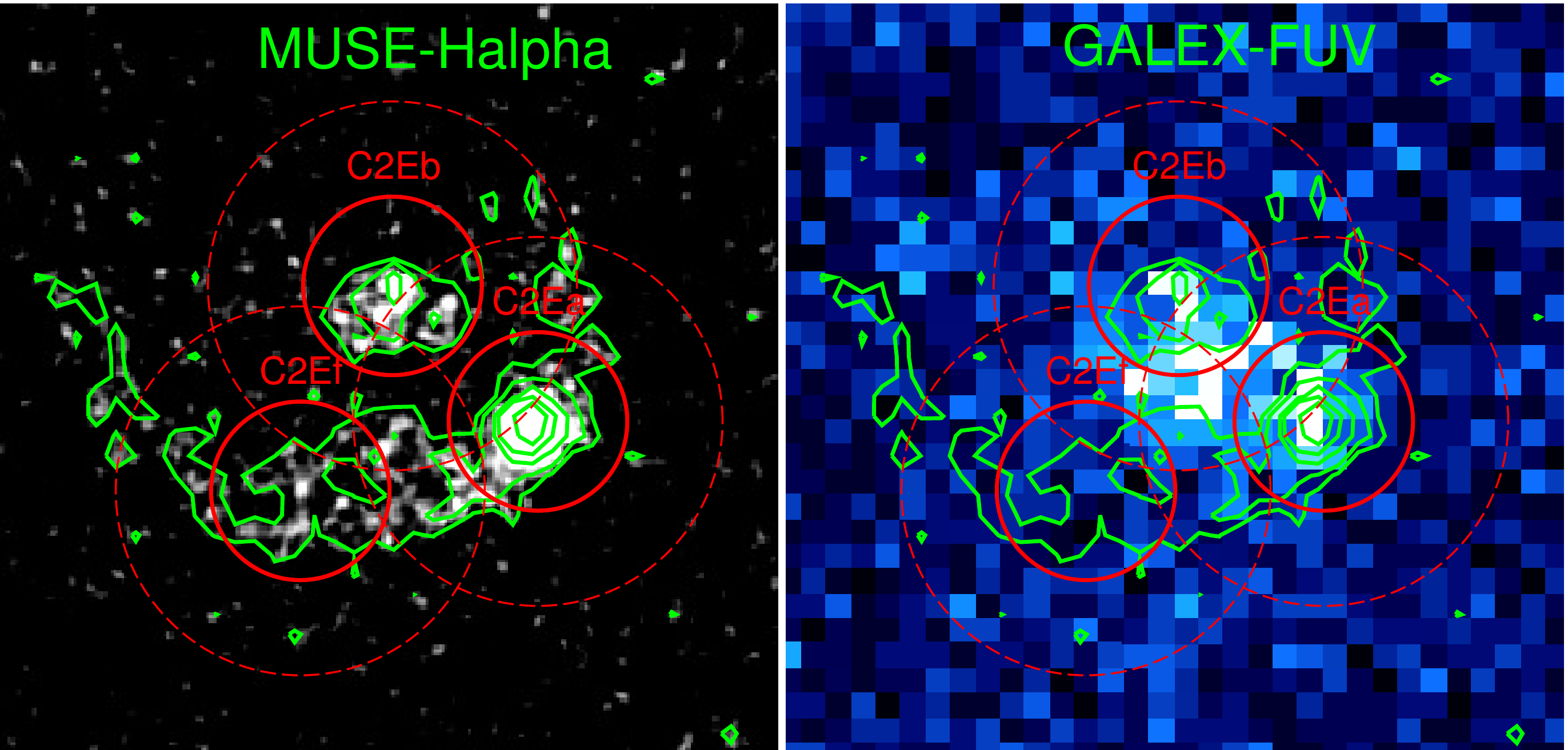}
 \caption{Contours of the H$\alpha$ emission of the two ionized regions in Clump2E, and of the partial ring  are overlaid in green on the MUSE-H$\alpha$ image in the left panel and on the GALEX-FUV continuum image in the right panel. Contour levels are: 1.3,2.5,5,10,25$\times$10$^{-20}$~erg~s$^{-1}$~cm$^{-2}$ per pixel (0.2\arcsec ).  Red continuous lines indicate the size of the  230~GHz FWHM IRAM-30m telescope beam  at the observed locations in Clump2E. Dashed red lines show the corresponding FWHM of the 115~GHz  beam. Observed position are labelled  just outside the continuous red line circles.  
}
\label{fig_c2e}
\end{figure*} 

The value of $\sigma$ has been computed in the heliocentric velocity  range  800 -- 1200~km~s$^{-1}$. A robust detection would show peak intensity above 5$\sigma$ in one of the lines with the second line at least marginally detected with peak or integrated line brightness above 3$\sigma$.  We don't have any robust detection but only a few marginal detections. Given the velocities of the H$\alpha$ lines we consider as marginal detections CO lines that lie in the velocity range  900 -- 1100~km~s$^{-1}$ and obey one of the following criteria:

-the peak intensity or the integrated line brightness is above 3$\sigma$ and lie within $\pm 20$~km~s$^{-1}$ of the mean H$\alpha$ velocity of the nearest HII regions or HI peaks;

-the peak intensity or the integrated line brightness is above 3$\sigma$ for both the CO J=1-0 and J=2-1 line; 

-the peak intensity or the integrated line brightness is above 4$\sigma$ for only one of the lines.

\begin{table*}
\caption{Parameters of the marginally detected $^{12}$CO lines} 
\centering                                       
\resizebox{19cm}{!}{%
\begin{tabular}{c c c c c c c c c c  }           
\hline\hline 
 Source & $^{12}$CO    & $\delta$V       &  $\sigma$ & I$_{sum}$        & I$_{gau}$  &  T$^{peak}_{gau}$   &   W$_{gau}$         & V$^{CO}_{hel}$ &V$^{H\alpha}_{hel}$ \\  
               &      line           &  km~s$^{-1}$ &  mK            &mK~km~s$^{-1}$ & mK~km~s$^{-1}$ &              mK        &   km~s$^{-1}$  & km~s$^{-1}$      & km~s$^{-1}$              \\  
 \hline\hline 
C1a&   J=1-0 & 0.5 & 1.9 & 7.0 $\pm$ 1.9 & 6.9$\pm$2      &  5.6    & 1.2$\pm$0.3 & 1000.0$\pm$0.2  & 994$\pm$2  \\
C1a&   J=2-1 & 0.5 & 1.9 & 10.1$\pm$2.9 & 10.6$\pm$3    &  4.3   &  2.3$\pm$0.3 & 1002.9$\pm$0.2   & 994$\pm$2 \\
C1b &  J=1-0 & 0.5 & 5.4 & 32$\pm8$       & 32$\pm$9      & 14.4   &  2.1$\pm$0.7   & 987.6$\pm$ 0.3   & 1003$\pm$3  \\
C1ab& J=1-0 & 0.5 & 1.7 & -8.6$\pm$1.8  & -6.2$\pm$1     &  -11.0   &  0.53$\pm$0.27 & 977.0$\pm$0.1     & 994-1003     \\
Cl2   & J=1-0 & 1.0 & 5.7 & 54$\pm$15    & 55$\pm$13    &  21.    &  2.4$\pm$0.7 & 1059.2$\pm$0.5   & 956(V$^{HI}_{hel}$)  \\
\hline\hline 
\end{tabular}
\label{lines}
}
\tablefoot{CO line units are main beam temperatures.}
\end{table*}

In Table~2 we show all lines that satisfy one of the above criteria and their estimated parameters. For marginal detections $\sigma$  has been computed  also in the 100~km~s$^{-1}$ interval around the line in the high resolution spectra ( 0.5~km~s$^{-1}$ channel width) and shown in Table~2. We have two marginal detections at the location of C1a and C1b, the two HII regions in Clump~1, whose spectra are shown in Figure~4.  In C1a, the line for the J=1-0 transition looks more promising than the higher level transition.  The remarkably low rms level of the C1a spectra has required 72~hours of telescope time. No J=2-1 counterpart is detected for the $^{12}$CO J=1-0 line marginally detected  in C1b and shown in Figure~5. For Cl2 J=1-0 line we find similar parameters  and integrated signal-to-noise  by choosing a spectral resolution   $\delta$v=0.5 or  $\delta$v=1.0~km~s$^{-1}$.  In Table~2  and in Figure~5 we show the results for the lowest spectral resolution for which we have the highest signal-to-noise for the line peak.  The peak velocity for Cl2 is however higher by about 100~km~s$^{-1}$ than the velocity of the HI peak. This, and the high value of $\sigma_{2-1}$, which does not allow to detect a possible J=2-1 line counterpart, casts some doubt on the effective presence of CO emission at the observed location. Within the 1'$\times$1' arcmin$^2$ MUSE field entered on Clump~2 no H$\alpha$ line has been detected. Unfortunately the FUV emission lies about 30~arcsec from the HI peak of Clump~2 and it is at the edge of the MUSE field, outside the FWHM of IRAM beams. Thus, we don't know if stars are currently forming in Clump~2. We underline that in other positions, like C1a and BSTD,  when the rms level in the averaged spectrum was much higher than the final one shown in Table~1, there were marginally detected lines that were not confirmed by longer integrations. Given the wide search range we consider the marginal detection in Cl2 as due to sporadic noise.

\begin{figure}
\includegraphics [width=10.2 cm, angle=0 ]{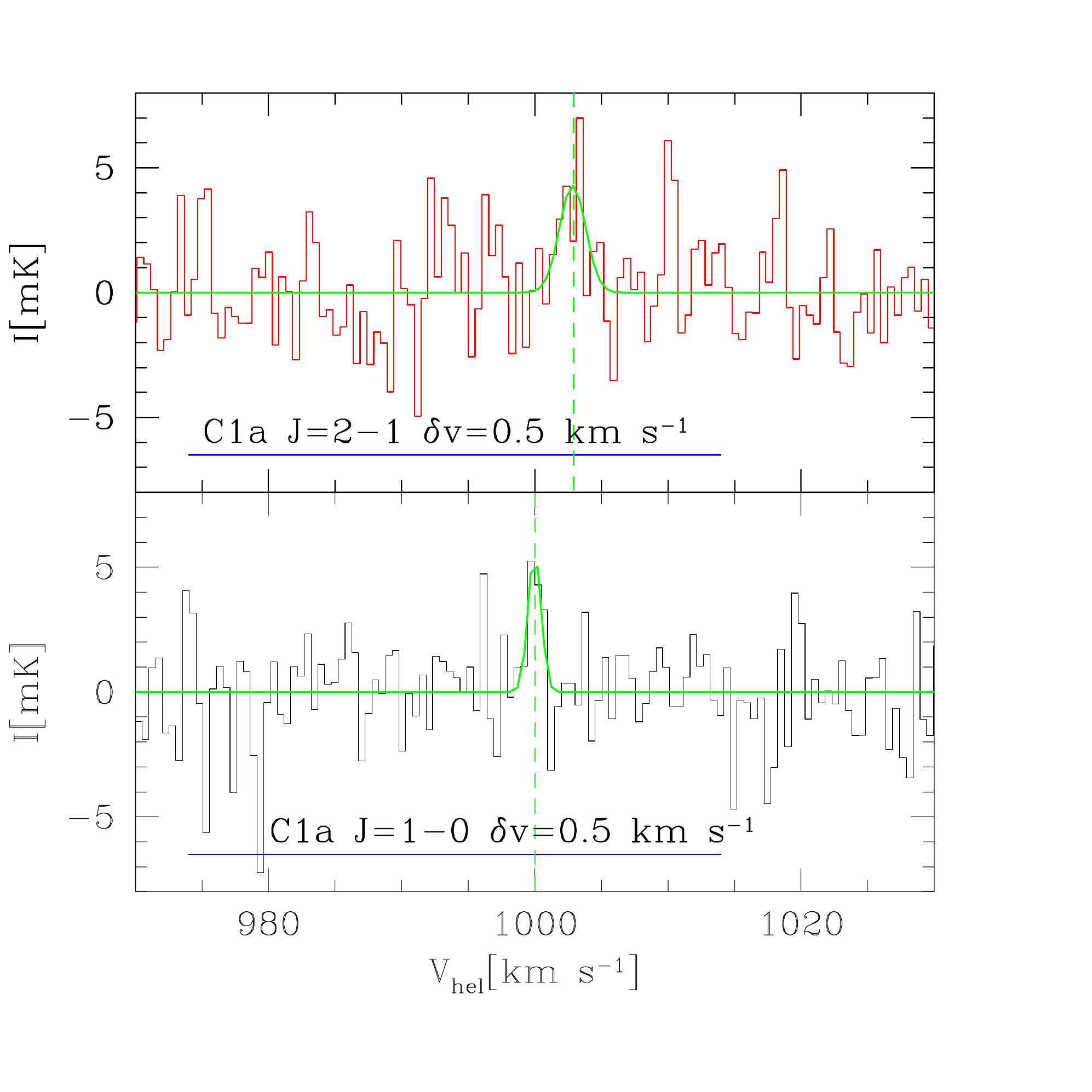}
 \caption{The CO J=1-0 and J=2-1 spectra at the position C1a are shown for a spectral resolution of 0.5~km~s$^{-1}$    in the bottom and upper panels respectively. Gaussian fits to lines marginally detected with the highest signal-to-noise integrated emission are shown with a green line. The velocities of the
 gaussian  peaks are marked by a green vertical  dashed lines. Horizontal blue lines at the bottom of each panel mark the expected velocity range (as given by the C1a H$\alpha$ line velocity $\pm$ 20~km~s$^{-1}$). }
\label{fig_c1a_sp}
\end{figure} 

\begin{figure}
\includegraphics [width=10.2 cm, angle=0 ]{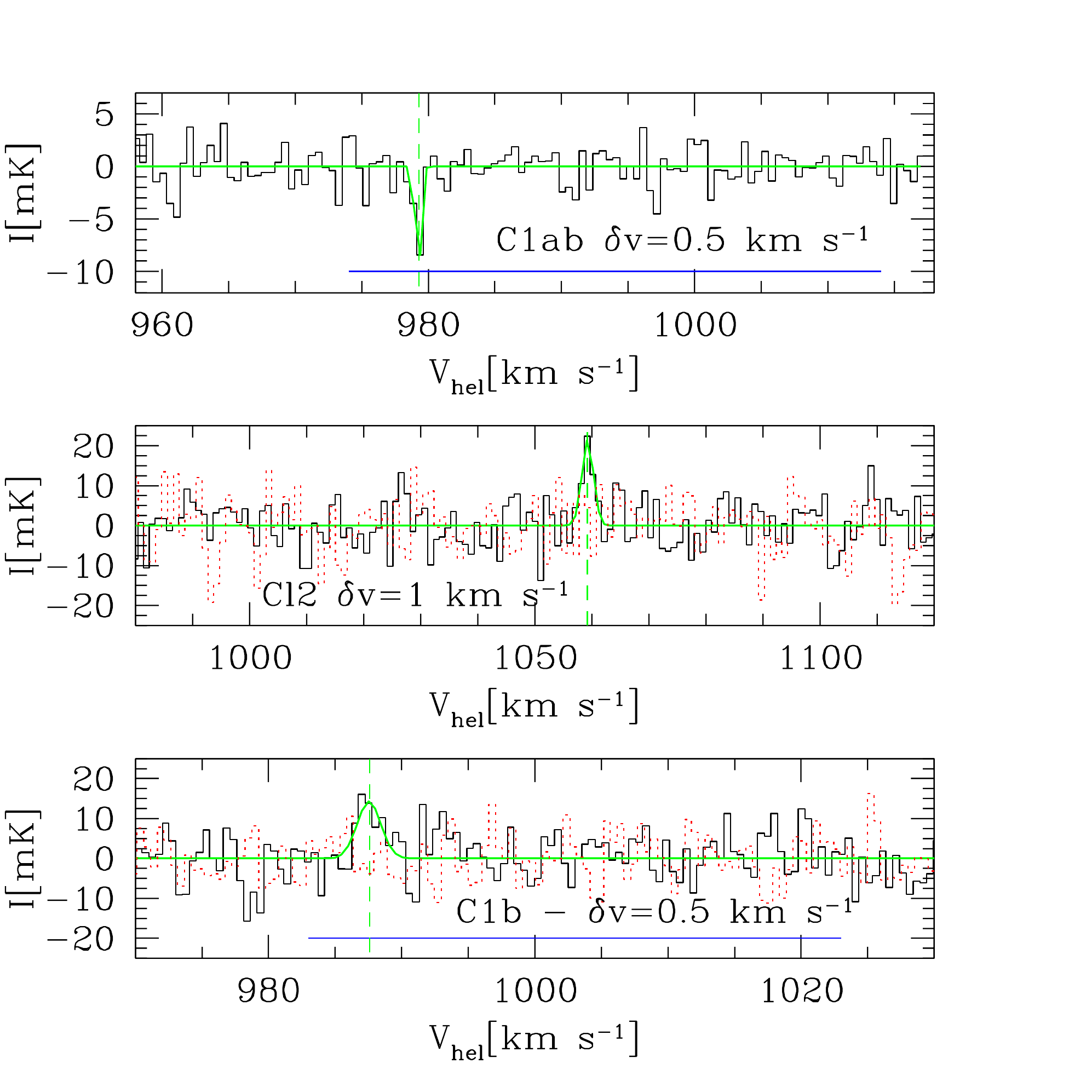}
 \caption{The CO J=1-0 spectra at the positions of C1b and Cl2  for  0.5 and 1~km~s$^{-1}$ channel width respectively are plotted in the bottom and middle panels.  The green lines show gaussian fits to marginal detections with the highest signal-to-noise integrated emission. Velocities of 
 gaussian  peaks are marked by  green vertical  dashed lines. The red dotted line shows the corresponding J=2-1 spectra. Horizontal blue lines at the bottom of each panel mark the expected velocity range (the Cl2 line is offset by more than 20~km~s$^{-1}$ from the velocity of the HI peak).  The top panel shows the CO J=1-0 stacked spectra of four adjacent positions in Clump1 (C1a, C1b, C1an, C1se). The gaussian fit to the narrow negative dip discussed in the text is shown by the continuous green line. }
\label{fig_c1a_sp}
\end{figure} 

Possible weak lines do not show up in the stacked spectra C1ab and this favors the hypothesis that the C1a and C1b lines are pseudo detections. The anomalous  negative narrow line is not detected in individual spectra, although the analysis of C1a and C1b spectra at high resolution underlines that a very narrow line is present at the same velocity in C1a during 2 of the 4 observing periods. At the same time, a wider but fainter negative feature is present in C1b spectra which might  enhance the C1a feature when stacked spectra are produced. No features are present in the same channel at the higher frequency and  this excludes a possible problem with a particular correlation channel. Having a FWHM of only 0.5~km~s$^{-1}$ (line width confirmed at higher spectral resolution) it is also very unlikely that the line is due to diffuse CO emission at the location of the wobble throw. Since the specific location of the reference spectrum changes with source elevation, we don't expect possible gas at the throw locations to have a coherent velocity. The feature is likely the result of a local interference.

\section{Are there candidate molecular clouds in Clump~1? }

In this Section we discuss the  marginally detected lines at the C1a and C1b locations, in the proximity of Clump~1  as shown in Table~2 and Figure~5. The spectral noise for these pointed  observations is very low and here a few very young massive stars are found.  All cloud models mentioned in this Section refer to the CO bright parts of GMCs and for these, given the close to solar metallicity measured (12+logO/H$\simeq$ 8.6, \citet{2021ApJ...908L..39C}), we  assume a standard solar CO-to-H$_2$ conversion factor X$_{CO}=2\times 10^{20}$~cm$^{-2}$~(K~km~s$^{-1}$)$^{-1}$. This converts the observed CO J=1-0 intensity I$_{CO}$, into  the beam diluted column density of molecular hydrogen, and it is equivalent to use the factor $\alpha_{CO}=4.3$~M$_\odot$ (K~km~s$^{-1}$pc$^2$)$^{-1}$ when inferring   the molecular mass density.  At the end of this Section we also discuss the possibility of CO-dark gas.

The star forming regions C1a  and C1b are close to the peak of Clump~1, the largest HI overdensity  in the main body of the ring at a mean heliocentric velocity of 987~km~s$^{-1}$ \citep{1986AJ.....91...13S}. Their separation from the nominal position of the HI peak is 27 and 19~arcsec. These are very rough estimates given the 45~arcsec FWHM of the VLA beam used for mapping the 21-cm emission \citep{1986AJ.....91...13S} . The nebula C1a is the second brightest HII region discovered by MUSE: the two CO line radial velocities  are in full agreement between them and with the velocities of the HI and HII gas (987 and 994~km~s$^{-1}$ respectively). For C1b we have an even better correspondence between the CO and the HI radial velocities (both at 987~km~s$^{-1}$).  We infer the  mass of molecular gas from the observed lines, its uncertainties and then discuss the possible physical characteristics of a population of molecular clouds in the area.

We can write the molecular mass, which takes into account the contribution of heavy elements,  as

\begin{equation}
 $
M$_{CO}$= 1.05$\times 10^4$  S$_{CO}^{1-0}$   D$^2_{Mpc}$ = 3.8$\times 10^3$  ${0.7 \over R_{21}}$ S$_{CO}^{2-1}$   D$^2_{Mpc}$$
\label{mcoeq}
\end{equation}

where S$_{CO}$ is the integrated line flux density in Jy~km~s$^{-1}$, R$_{21}$ is the intrinsic  (2-1)/(1-0) line ratio( i.e. in brightness temperature), and D$_{Mpc}$ is the distance to the source in Mpc \citep{2013ARA&A..51..207B}.
The observed integrated line intensity I$_{CO}$ measures the beam diluted brightness temperature T$_b$ $\Delta$v= I$_{CO}$ $\Omega_{s*b}$/$\Omega_s$ where $\Omega_{s*b}$ solid angle is the source convolved with the telescope beam and it is equal to the telescope solid angle when the source is much smaller than the beam.  We work under this assumption, and given the IRAM telescope parameters, we convert the intensity in K main beam temperature units into flux densities in Jy as

\begin{equation}
\frac{S_{CO}}{ [{\hbox {Jy~km~s}}^{-1} ]}=5\ \frac{I_{CO}} {[{\hbox{K~km~s}}^{-1} ]}
\end{equation}

Assuming that the emitting clouds in C1a and C1b are at the beam center, the observed CO line fluxes imply masses of   3.7$\times 10^4$~M$_\odot$ and 1.7$\times 10^5$~M$_\odot$ respectively i.e. a total mass of the order of 2$\times 10^5$~M$_\odot$. This is lower than that estimated from the standard value of the gas consumption time  discussed at the end of this Section.
 
 For sources that are offset with respect to the beam center, the molecular mass using  equation (\ref{mcoeq})   is underestimated. In M33 the spatial correlation between infrared selected young stellar clusters  and giant molecular clouds is strong, with a typical separation of 17~pc \citep{2017A&A...601A.146C}. This displacement, less than 1~arcsec at the distance of the Leo ring, would be negligible for the IRAM beams. Considering the nearest galaxies of the PHANGS sample, NGC628, NGC5068, NGC5194, observed  in the CO J=2-1 line with a good spatial resolution, similar to that of M33, the separation between independent star forming regions is of the order of 100~pc \citep{2020MNRAS.493.2872C}. Similarly the nearest neighbour separation between massive GMCs and young stellar clusters in a larger PHANGS sample is found to be of the order of 100~pc \citep{2022arXiv220902872T}. This suggests  that the separation between HII regions and the associated native molecular clouds, or the molecular cloud formed by  feedback, is expected to be of the order of  100~pc  for H$\alpha$ selected star forming regions. This separation is much smaller than the beam HWHW at 115~GHz. Corrections to the observed flux become severe when the separation between the emitting object and the beam center is larger than the beam HWHM.  Feedback effects might be stronger in the Leo ring than in galaxies and we can use the HII region radius,  of the order of 150~pc (3~arcsec), as a rough estimate for expected cloud-HII region  separation. Considering projection effects and pointing errors, of the order of 1.5~arcsec, the  cloud-beam center offsets are expected to be less than  6~arcsec. This can reduce up to about 50$\%$ the flux recovered for the CO J=2-1 line but  the effects on the CO J=1-0 line are much smaller.  

\subsection{Relating cloud physical parameters}

To estimate the cloud radii in parsecs, R$_{pc}$, we use the viral theorem assuming a cloud density profile and  a viral to CO luminous mass ratio. For  $\rho \propto r^{-1}$ we have 

\begin{equation}
\frac{M_{vir}}{M_\odot}= 188\  {\hbox {R}}_{pc}\times {\hbox {FWHM}}^2 
\label{vir}
\end{equation}

where the FWHM is in km~s$^{-1}$ \citep{2013ARA&A..51..207B}. Assuming that  M$_{vir}$/M$_{CO}$=1.0, as it is often observed in galaxies  \citep{2013ARA&A..51..207B}, and using the FWHM of the marginal detections we infer very large cloud radii (136 and 206~pc  for C1a and C1b respectively). 
These  radii, given the cloud mass,  imply extremely low molecular gas volume densities, which are unphysical given the inefficient formation of molecules when gas densities are below 30~cm$^{-3}$ \citep{2022ApJ...931...28H}. 
In addition, the marginally detected lines  are narrow for their observed brightness, and not compatible with the standard size-linewidth relation which reads \citep{2013ARA&A..51..207B}:

\begin{equation}
{\hbox {R}}_{pc}  = C_{SL}\ {\hbox {FWHM}}^2  
\label{slr}
\end{equation}

with $C_{SL}=0.37$~pc~(km~s$^{-1}$)$^{-2}$ and which implies parsec size clouds for the observed linewidths.  The standard size-linewidth relation predicts  cloud masses that scale with radius or FWHM as follows:

\begin{equation}
\frac{M_{vir}}{M_\odot}=188\  C_{SL}\ {\hbox {FWHM}}^4 = 188\frac{{\hbox {R}}_{pc}^2}{C_{SL}}
\label{virial}
\end{equation}

\noindent
In this case the cloud mass surface density  is :

\begin{equation}
\frac{\Sigma_{mol}} {M_\odot pc^{-2} }=\frac{M_{CO}}{\pi {\hbox {R}}^2_{pc}}=\frac{188}{ \pi {\hbox {C}}_{SL}}
\end{equation}

This is constant and of the order of 162~M$_\odot$~pc$^{-2}$  \citep{2013ARA&A..51..207B},  much higher than the 0.6~M$_\odot$~pc$^{-2}$ we estimate at C1a. Any additional mass, such as CO dark mass in a virialized cloud core or any mass corrections for cloud offsets with respect to the beam center,  will increase the deviations from the size-linewidth relation observed in galaxies. 

\subsection{The standard cloud model}

The CO lines shown in Table~2 are tentative detections with large uncertainties.  In this subsection we assume that  lines listed in Table~2 are not true detections but false signals driven by sporadic noise. In this case by considering clouds with physical parameters similar to those in galaxies,  undetected because their emission is below the detection threshold, we can derive upper limits to their masses. The data on the J=2-1 line are not used in this Section  because of uncertainties  in the intrinsic line ratio (gas densities might not be sufficiently high that collisional equilibrium is reached for this higher level line), and because of possible cloud displacements with respect to the smaller 230~GHz beam.  The J=2-1 data will be examined in the next Section when we consider the beam diluted molecular gas surface brightness. 

\begin{figure*}
\includegraphics [width=9.0 cm, angle=0 ]{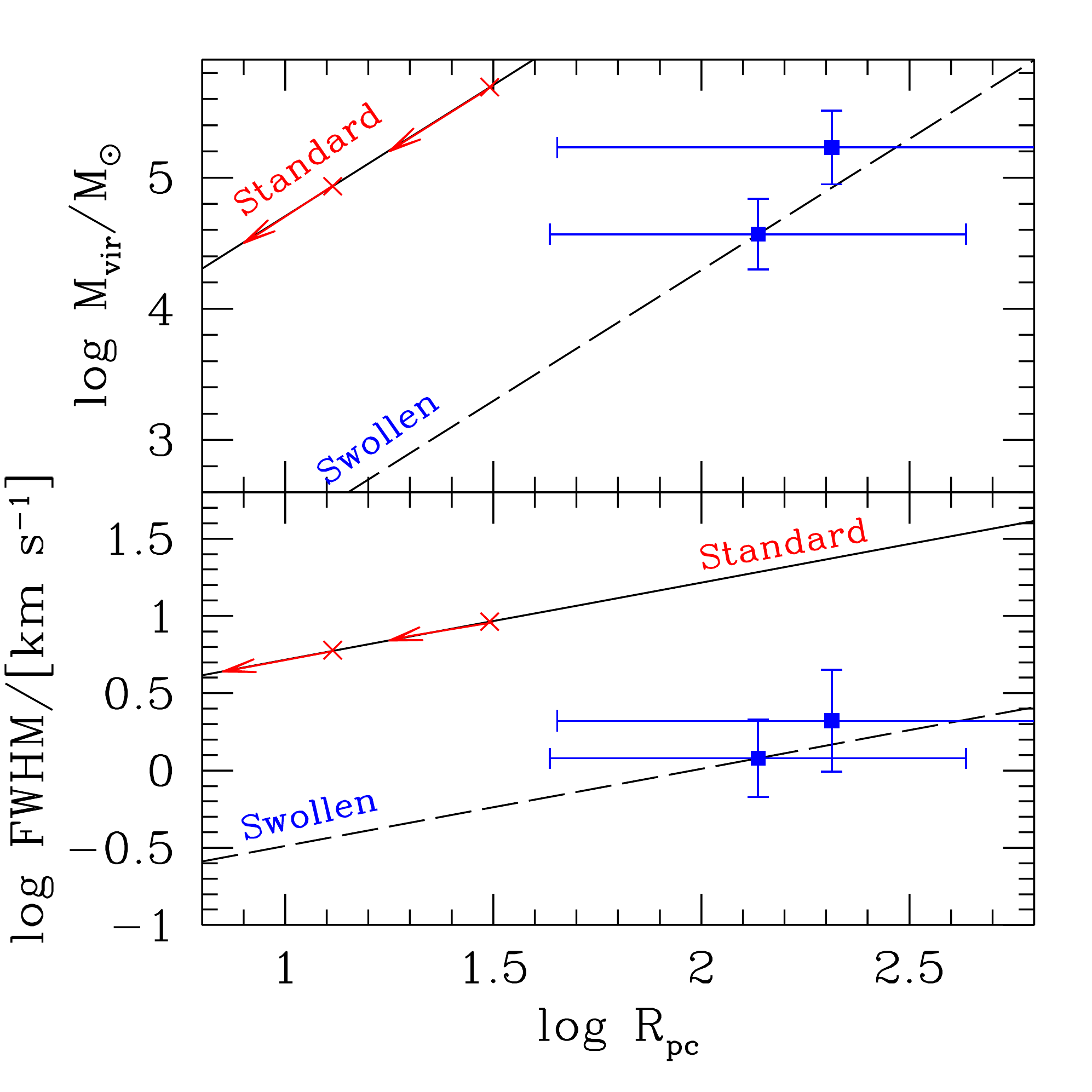}
\includegraphics [width=9.0 cm, angle=0 ]{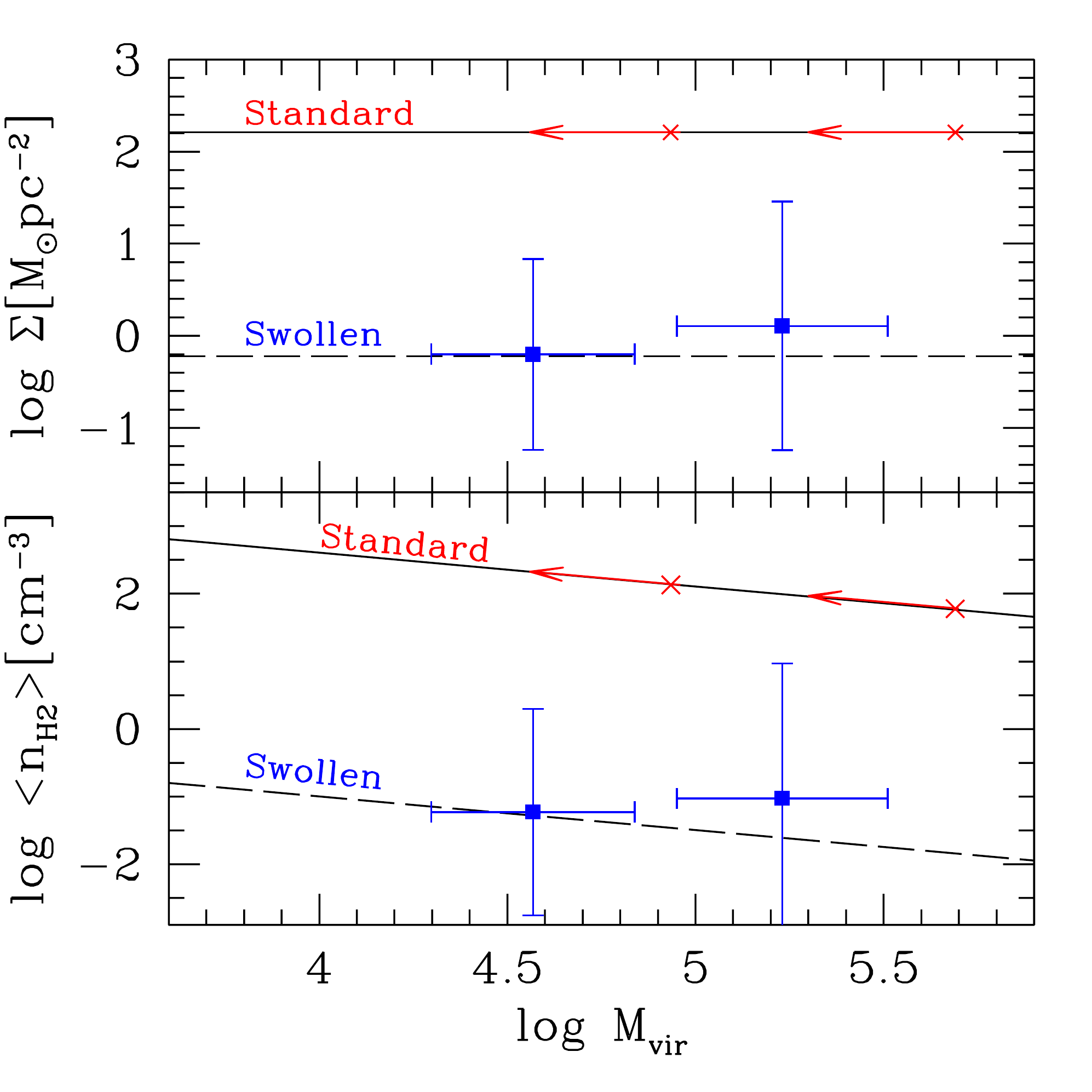}
 \caption{ In the left panels we show how the cloud viral radius relates to the line FWHM (bottom panel) and to the cloud viral mass (upper panel) according to the {\it standard} cloud model. The long dashed line is for the {\it swollen} model which reproduces the tentative detections. Similarly in the right panels  we show the relation between the cloud viral mass and the  average   volume density (bottom panel)  or mass surface density (upper panel) for the {\it standard} and the {\it swollen} model. The tentative detections for C1a and C1b are shown with blue triangles for the {\it swollen} model. For the {\it standard} model we only have upper limits which we show with red arrows following the cross symbols. 
 }
\label{mod}
\end{figure*} 

For the standard model  we assume that the size-linewidth relation observed in galaxies and described by equation (\ref{slr})  applies. 
If a population of clouds with cloud masses lower than M$_{max}$ is in place we can compute M$_{max}$ from the rms of the observed spectra. We focus here on C1a and C1b areas. We compute the maximum mass as a function of the J=1-0 spectral rms $\sigma_{mK}$ in mK at a spectral resolution $\delta$v in km~s$^{-1}$ as follows:

\begin{equation}
\frac{M_{max}}{M_\odot}=\frac { 5\times 1.05\times 10^3\times 3\sigma_{mK}\times {\hbox {FWHM}}} { \sqrt{\hbox {FWHM}/{\delta v}} }
\label{max1}
\end{equation}

\noindent
setting this maximum mass equal to  M$_{vir}$  we have that 

\begin{equation}
\frac{{\hbox{FWHM}}}{km~s^{-1}}=\Big(\frac{M_{vir}}{188C_{SL}}\Big)^{0.25}=\frac{M_{max}^{0.25}}{2.9}
\label{max2}
\end{equation}

Substituting this expression for the FWHM into eq.(\ref{max1}) for $\delta$v=2~km~s$^{-1}$ we have M$_{max}$ as a function of the values of $\sigma_{mK}$ quoted in Table~1, that reads:

\begin{equation}
\frac{M_{max}}{M_\odot}=8\times 10^4 \sigma_{mK}^{1.14}
\label{max3}
\end{equation}

The lowest value of  M$_{max}$ is 9$\times 10^4$~M$_\odot$ for C1a, where we have the lowest noise, the highest value of M$_{max}$ is 5$\times 10^5$~M$_\odot$ at C1an. The upper limit M$_{max}$ in C1a is larger than the tentatively detected mass  because of  the wider line  considered.
If clouds are not at the beam center but within the beam FWHM, a slightly higher maximum mass should be considered. 

In the left panel of Figure \ref{mod} we show the FWHM and M$_{vir}$ as a function of cloud radius R$_{pc}$ for the  standard model considered. Upper limits to C1a and C1b are shown with red crosses. For comparison, we also plot with dashed lines a scaled version of Eq.~\ref{slr}, where a different coefficient than C$_{SL}$ has been used and determined by fitting the marginally detected CO J=1-0 line in C1a. Although this model, referred to as the swollen model, is in agreement with the pseudo-lines  in the spectra of C1a and C1b (shown with filled square symbols)  we will see that it predicts too low densities for the gas to be molecular.  In the right panel of the same Figure we show  in fact the average molecular hydrogen number density  and the molecular mass surface density in the clouds (computed as ratio between the cloud mass M$_{CO}$ and the cloud virial volume or area) for the same models as a function of the cloud viral mass. 

\subsection{CO-dark gas and the gas self-shielding conditions}

It is clear from Figure \ref{mod} that the marginally detected lines predict three orders of magnitude lower mean cloud densities than standard clouds observed in galaxies.
This cast some doubt on the gas ability to form molecules and to self-shield them from UV radiation from the nearby massive stars. If UV radiation from an  HII region or from the interstellar field is incident on a neutral cloud, H$_2$ molecules can  reach high fractional abundance only  after one magnitude of visual extinction, while CO is found deeper into  denser clumps, where A$_V > 2$~mag  or N$_H >4\times 10^{21}$~cm$^{-2}$ \citep{1999RvMP...71..173H}.  
Estimates of the visual extinction, made through the Balmer decrement observed in optical spectra, give A$_V\sim 0.5$  \citep{2021A&A...651A..77C}. The presence of  CO-dark envelopes can increase the cloud column density and attenuate the UV radiation field. Numerical simulations have  shown that these envelopes are very common when gas volume densities are low, even for metal abundances close to solar  \citep{2022ApJ...931...28H}. However, CO-dark envelopes lower the SFE but do not help solve the problem of the low number gas densities in CO bright clumps. The low gas densities suggest low factional abundances of molecular hydrogen  and long free-fall times.

Theoretically the detailed formation-dissociation balance for H$_2$ and CO molecules depends on the ratio n/G where n is the gas volume density and G is the intensity of the UV radiation field. The radii of  C1a and C1b HII regions are about 200 and 100~pc respectively. Given the stellar mass and age of their associated stellar clusters \citep{2021A&A...651A..77C}, we estimate a UV interstellar radiation field at the edge of the HII regions that is between 10 and 5 times lower than in our local interstellar medium.  As n/G decreases the molecule's fractional abundances decrease. If a few small dense clumps  are floating in a large more massive diffuse cloud, mostly atomic,  the whole cloud might show a low mean molecular fraction (if pressure supported or self-gravitating). However, the marginally detected lines can hardly refer to clumps floating in a diffuse clouds  given the 10~km~s$^{-1}$ FWHM for HI lines usually observed in the ring  \citep{1986AJ.....91...13S} although high resolution HI data are needed to verify that these large  FWHM apply also to hundred parsec scale regions. 

The considerations expressed in this subsection suggest 
 that the marginally detected lines  should be considered with extreme caution and are likely due to sporadic noise.  Future investigations through  HI and CO observations at high spatial resolution, as well as specific theoretical models,  are needed to test the turbulence level and other properties of dense gas fragments that form the sparse populations of stars in the Leo ring.

\section{The star formation law at its lowest extreme}

The relevance of CO detections in the Leo ring is not only related to understanding the physics and chemistry of cold gas overdensities  forming stars in quiescent intergalactic clouds  but also to the fact that any detection  might represent the lowest extreme ever observed for the molecular Kennicutt-Schmidt law \citep{2012ARA&A..50..531K}. It  might confirm or refute the universality of such relation when the star formation rate surface density is lower than in outer disks or gas is not rotationally supported. We test the molecular Kennicutt-Schmidt law considering  upper limits for the CO lines and the standard cloud model for all  IRAM-30m pointed observations of star forming regions in the Leo ring. We  note that results are similar if future observations should confirm  the marginally detected lines. We compare  the gas consumption time and the star formation efficiency determined in disk galaxies to estimates for the Leo ring star forming regions.

\subsection{ The molecular Kennicutt-Schmidt relation}

We consider the six  pointed observations listed in Table~1 which cover star forming regions imaged by MUSE and have $\sigma_{1-0,2-1}\le 5$: C1a, C1b, C1ase, C2Ea, C2Eb, C2ef. For these regions we perform circular aperture photometry in H$\alpha$ to measure the SFR and its surface density, $\Sigma_{SFR}$. We choose two apertures, with the same angular size as the IRAM beam solid angle at 115~GHz and at 230~GHz (circular apertures with radii of 6.6 and 13.2~arcsec  respectively). For the larger beam the emission in C1a and C1b is the same at both locations and it gives a SFR of 8$\times 10^{-4}$~M$_\odot$~yr$^{-1}$  using the conversion factor in Table~3 of \citet{2021A&A...651A..77C} for M$_{up}$=35~M$_\odot$ which reads

\begin{equation}
{{ SFR_{H\alpha}}\over [M_\odot~{\rm yr}^{-1}]} = 1.9 10^{-41}{L_{H\alpha}\over [{\rm erg~s}^{-1}]} 
\end{equation}

The limiting mass for the upper end of the IMF has been determined by \citet{2021A&A...651A..77C} by comparing optical data with simulated stellar cluster models. Similar upper mass limits have been derived for the young population in outer disks \citep{2020MNRAS.491.2366B}. We correct for an average visual extinction A$_V$=0.5. Uncertainties on the SFR are of the order of 0.5~dex, mostly driven by stochasticity due to incomplete sampling at the high mass end of the IMF   \citep{2021A&A...651A..77C}.  The star formation rate density  in C1a and C1b is 0.6$\times 10^{-4}$~M$_\odot$~yr$^{-1}$~kpc$^{-2}$ in the 115~GHz beam. We compute the star formation rate density for all the star forming regions listed in Table~1.

To evaluate the molecular mass in the same aperture, we use the data in  Table~1.  We assume  an intrinsic line ratio  R=2-1/ 1-0 = 0.5 and use also the data of  the J=2-1 spectra. 
We compute the molecular mass surface density in the beam assuming that the source is much smaller than the beam as 

\begin{equation}
\Sigma_{mol}=\Sigma_{CO}\frac{\Omega_s}{\Omega_{b*s}}=\frac{M_{CO}}{\Omega_b}
\label{sb}
\end{equation}

\noindent
where $\Sigma_{CO}$ is the cloud column density and M$_{CO}$ is the cloud mass inferred from CO integrated emission line. The beam solid angle $\Omega_b$ is 1.2 10$^6$ and 3 10$^5$~pc$^2$ for the IRAM beam at 115~ and 230~GHz respectively at the distance of the Leo ring. The expression in equation (\ref{sb}) is equivalent to multiply directly the integrated main beam temperature  by the CO-to-H$_2$ conversion factor i.e. it measures the mean mass surface density also in the case of a population of low mass clouds in the beam.  

 \begin{figure*}
\includegraphics [width=16 cm, angle=0 ]{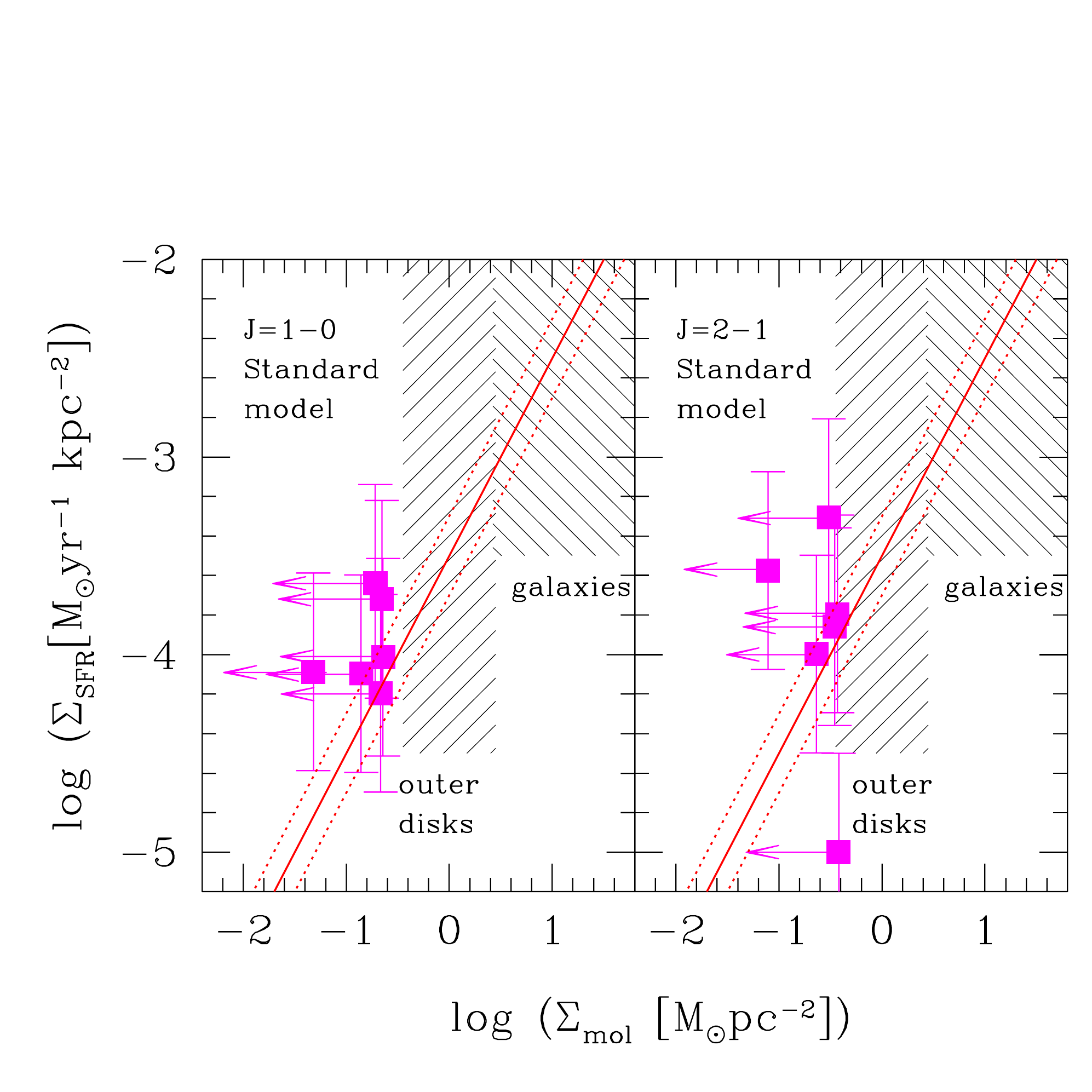}
 \caption{
The magenta square symbols indicate the upper limits to the average molecular mass surface densities in the IRAM beams for all the 6 star forming regions  considering the standard cloud model for which the FWHM of the line is computed consistently and marginal detections are considered spurious noise.   The continuum and dotted lines indicate the K-S relation and its dispersion as fitted by \citet{2013AJ....146...19L} to inner disk data of galaxies. The shaded areas indicate the molecular gas surface densities typical of the inner regions of galaxies ("galaxies" label) and of outer disks (see \citet{2016MNRAS.455.1807W}).    }
\label{ks2}
\end{figure*} 

We use the  standard cloud model  to derive the limiting mass surface densities for all  6 regions as in eq.(\ref{max1}), considering C1a and C1b lines as spurious noise and the standard cloud model to infer  FWHM$_{lim}$. This is the FWHM of the most massive cloud compatible with the data  which is computed  following equation (\ref{max1}) and equation (\ref{max2}):

\begin{equation}
M_{CO,lim}=188\  C_{SL}\  FWHM_{lim}^4 = 1.58\times 10^4 \sigma_{mK} \sqrt{FWHM_{lim}\ \delta v} 
\end{equation}

\noindent
from these equations we find FWHM$_{lim}$ and the associated limiting mass

\begin{equation}
FWHM_{lim}=\frac{4\times \sigma_{mK}^{0.29}}{C_{SL}^{0.29}} \Rightarrow M_{CO,lim}=4.5\times 10^4 \frac{\sigma_{mK}^{1.14}}{C_{SL}^{0.14}}
\label{mlim}
\end{equation}

\noindent
The line FWHM in this case  is generally smaller than 10~km~s$^{-1}$ varying  between 5.3 and 8.1~km~s$^{-1}$ for the six star forming regions considered in this paper.  
 We  compute the upper limits to the average molecular mass surface densities in the IRAM beams, for these six regions. We use magenta filled symbols  and plot these values in the Kennicutt-Schmidt diagram in Figure~(\ref{ks2}). The solid straight line shown fits the data of the molecular mass surface densities of galaxies as given by \citet{2013AJ....146...19L}.  Dotted lines show the dispersion around the fitted line. At lower densities we mark the range of $\Sigma_{mol}$ for outer disks according to the data collected by \citet{2016MNRAS.455.1807W}. The Leo ring data are at extreme low values of molecular gas surface densities and star formation rate densities, where the K-S relation has not been tested yet. A similar relation for the atomic gas surface densities has been shown by \citet{2021A&A...651A..77C} and implies extremely long depletion times for the HI gas.
 
Figure~\ref{ks2} indicates that most of the regions lie above the K-S relation. In particular the location of C1a (the leftmost star forming region) and that of C2Ea (the region with the highest SFR density) in Figure~(\ref{ks2}) seem incompatible with an extrapolation of the molecular K-S law towards low density regions. This inconsistency, evident both for the J=1-0 line and for the J=2-1 line, is then independent from the cloud model considered. The beam area for the J=1-0 line is of the order of 1.2~kpc$^2$, fifty times larger than the size of the HII regions, and of the same order of the scale sampled by \citet{2013AJ....146...19L}. The surface density of molecular gas is lower than expected and implies  a very short molecular gas depletion time, opposite to the atomic gas depletion time.

 \subsection{Star formation efficiency and gas consumption times}

The average gas depletion time,  DT, and the star formation efficiency, SFE, are usually defined and related to each other as follows:

\begin{equation}
DT= M_{mol}/SFR \qquad SFE=M_*/M_{mol} = \tau_{GMC}/DT
\end{equation}

where M$_*$ is the stellar mass of the young stellar cluster born in a molecular cloud of mass M$_{mol}$, and $\tau_{GMC}$ is the active lifetime of molecular clouds. Sometime the star formation efficiency per free-fall time is used by substituting $\tau_{GMC}$ with the free-fall time at a given spatial resolution.

The average gas depletion time is  observed to be relatively constant in nearby spiral  and tidal dwarf galaxies  and equal to 10$^{9.2}$~yr,\citep{2011ApJ...730L..13B,2011AJ....142...37S,2017A&A...607A.110L}. Given the  SFR  measured by \citet{2021A&A...651A..77C}, for each region considered in this paper,  a DT similar to that observed in disk galaxies implies molecular masses in the range 0.4--4$\times 10^6$~M$_\odot$. Our most sensitive observations  show that these masses are  overestimated. For the case of C1a the depletion time is of the order of 3$\times 10^8$~yr,  a factor 5 lower than in galaxies.  Similar conclusions hold if molecular clouds are small but widespread as in outer disks \citep{2017A&A...601A.146C,2019A&A...622A.171C}. In this case  we expected an average molecular gas surface density  of order 0.5~M$_\odot$~pc$^{-2}$ \citep{2015ARA&A..53..583H},  higher than observed in the Leo ring, not compatible with the most sensitive data analyzed in this paper. 

We  can compute the star formation efficiency, SFE, as the ratio between the stellar mass of young clusters and the mass of the nearest molecular cloud i.e. assuming that the cluster progenitor cloud has a similar mass to other clouds in the cluster proximity.  The estimated cluster masses at C1a and C1b are in the range 500-1000~M$_\odot$ and we expect progenitor clouds  to have masses in the range 5-10$\times 10^4$~M$_\odot$ for a typical star formation efficiency of 1$\%$. Therefore clouds similar to those progenitors  should have been detected  by the deep integration performed in the C1a, even for the standard cloud model.  Lower M$_{mol}$ imply a higher SFE. 

Cloud active lifetimes of the order of 10-20 Myr have been measured by associating molecular clouds detected in deep survey of nearby galaxies to embedded (only infrared radiation detected) and exposed (with detectable H$\alpha$ or optical continuum emission) young stellar clusters  and by determining the age of the exposed ones \citep{2009ApJS..184....1K,2017A&A...601A.146C}.  
A 10~Myr active lifetime and the measured DT give a SFE of order of 3$\%$, only slightly higher than determined by observations and  theoretical models of bright spirals a \citep{2013MNRAS.432..653D,2018ApJ...861L..18U}. A somewhat shorter cloud lifetime is more in agreement with standard SFE values.

We can therefore  conclude, using the data for the young and more active star forming regions of the Leo ring,  that molecular clouds form stars for a shorter period  in the ring  low density environment than in disk galaxies. The shorter star formation timescale indicates that clouds might be easier to dissolve as stellar feedback switches on and, more in general, that the gas-to-star conversion process in the Leo ring might depart from that studied extensively in disk galaxies. This might explain the nature of ultra diffuse dwarf galaxies. 
If molecular clouds dissolve quickly,  or have a very high efficiency in forming stars, they will hardly be detected close to young stellar clusters. 
Our analysis shows that the decrease of the molecular gas depletion time observed from spiral to dwarf galaxies might not only be due to a decrease of the gas metallicity   but also to variations of the local conditions such as a decrease of stellar mass and angular momentum, as previously claimed  \citep {2011MNRAS.415...61S, 2013AJ....146...19L, 2015A&A...583A.114H}. The possibility of CO dark gas  in a low density but  solar metallicity environment, such as the Leo ring, needs to be investigated in more detail and it might be linked to a change in the cloud's physical conditions. 
The formation of stars from atomic gas should  also be considered in case future more sensitive observations  do not confirm the presence of  molecular clouds   \citep{2012ApJ...759....9K,2012MNRAS.421....9G}. Sparse CO bright dense clumps in extended self gravitating atomic envelopes might be progenitors of  the sparse population of young stars  forming  in Clumps~1.

 \section{Summary and conclusions}

The close to solar metallicity recently determined for the giant HI ring in Leo has solved the mystery of the ring origin but at the same time the discovery of a  sparse population of massive young stars has triggered new wonders. Why has the gas in this collisional ring  not been forming stars since it has been stripped from its progenitors? And how are stars  formed in an environment more quiescent than a galactic disk? 
how massive are molecular clouds and how efficiently  are they  forming stars?  Moreover, the question of whether  stars are born in clusters, associations or in isolation in such  an environment  still needs to be investigated.
 
In this paper we have presented the results of deep pointed observations of CO J=1-0 and J=2-1 lines in recently discovered star forming regions of the Leo ring and close to its two most prominent HI peaks. These observations have reached a spectral sensitivity that is  between 27 and 6 times better than previous observations of the CO J=1-0 line   \citep{1989AJ.....97..666S}. The new data sensitivity is sufficiently high to possibly detect the  expected beam diluted molecular gas surface density,  given the  observed star formation rate density. The relation between the gas mass surface density  and the star formation rate density has never been tested for  such extremely low densities,  as observed in the Leo ring.  

For 11 pointed observations we reached a spectral noise between 1 and 5~mK main beam temperature, at least in one  observed frequency when spectra are smoothed at  2~km~s$^{-1}$ resolution. The most sensitive observations have been carried out for C1a, a region where  stars are forming at a rate  of the order of 10$^{-4}$~M$_\odot$~yr$^{-1}$~kpc$^{-2}$. The average mass surface density of molecular hydrogen in the 115~GHz beam expected from the extrapolation of the observed K-S relation  in resolved star forming galaxies, should be higher than 0.1~M$_\odot$~pc$^{-2}$.  The marginally detected line gives instead a lower mass surface density, only 0.03~M$_\odot$~pc$^{-2}$. If this marginal detection is spurious noise we derive a similar upper limit: 0.04~M$_\odot$~pc$^{-2}$. For the most active star forming region found in the ring, C2Ea, the star formation rate density is 5$\times 10^{-4}$~M$_\odot$~yr$^{-1}$~kpc$^{-2}$ in the 230~GHz beam, but the surface density of molecular hydrogen is more than three times lower than expected. 

The IRAM-30m  sensitive observations imply that the physical parameters of the star formation process in the Leo ring  might deviates from those established in disk galaxies. We summarize below  a few scenarios, that might explain deviations  from the K-S law:

\begin{itemize}
\item
The molecular gas depletion timescale is short  or, equivalently, gas condensations are very efficient in forming stars in the Leo ring.  
The gas in giant molecular clouds is more prone to rapidly dissolve or the low star formation rate and the lack of rotation  might  favor the formation of  only low mass clouds with a high SFE.
\item
The mass of CO bright clumps might be only a fraction of the  cloud mass that collapses and fragments to make stars. This implies  CO dark molecular gas or that  collapsing gas is not fully molecular. 
\item
The mean separation between HII regions and the nearest giant molecular clouds might larger than 500~pc because of efficient stellar feedback; molecular clouds in this case escaped detection because of the limited spatial coverage of IRAM-30m pointed observations.
 \item
Lastly, the stellar IMF might deviate from the power law observed in galaxies by suppressing gas fragmentation, or gas collapse, at small masses. The decrease of the estimated star formation rate makes the gas depletion timescale longer. Although this runs counter to generally considered top-light IMFs for low mass stellar clusters, only deep fully-resolved stellar color-magnitude diagrams  could reduce this uncertainty through explicit star counting and mass estimation.  

\end{itemize}

\noindent
These conclusions hold regardless of whether the marginally detected lines in  C1a and C1b are due to emission from swollen molecular clouds or  to spurious noise. We underline that the very narrow and uncertain line widths, the low signal-to-noise  ratio and the predicted low gas number densities cast doubt on the identification of the marginally detected lines as true line emission.
For this reason, we have focused our analysis considering upper limits to the CO emission consistent with the standard size-linewidth relation that holds for CO bright virialized clouds in galaxies.

Our results indicate that the lifecycle of molecular clouds in the Leo ring might be shorter than in disk galaxies or, equivalently, the spatial offsets between a young stellar cluster and a new ongoing  embedded star formation site are much larger than in disk galaxies. This  might explain the nature of ultra diffuse dwarf galaxies. Future sensitive observations at higher spatial resolution with mm interferometers, such as ALMA, together with additional data from  21-cm and Hubble Space Telescope imaging, can help  us distinguish if  deviations of  the Leo ring star formation law  from the K-S relation are due to  variations in the mass distributions of gravitationally bound molecular clouds or to variations in their physical characteristics. The abundant J=3-2 CO detections  in the outer disk   of M83 \citep{2022arXiv220811738K} suggest that other CO transitions may further clarify the physical state of the molecular ISM in extreme environments.

\begin{acknowledgements}
We thank the IRAM allocation time committee and IRAM-30mt operators during lockdown period for their kind assistance during remote observing.
\end{acknowledgements}

 
 \end{document}